\documentclass[twocolumn,showpacs,preprintnumbers,amsmath,amssymb]{revtex4}

\usepackage{graphicx}
\usepackage{dcolumn}
\usepackage{bm}

\begin{document}

\preprint{APS/123-QED}

\title{Mori-Zwanzig projection formalism: from linear to nonlinear}

\author{Jianhua Xing}
\homepage{http://www.biol.vt.edu/faculty/xing}
 \email{jxing@vt.edu}
\affiliation{%
Department of biological sciences, Virginia Polytechnic Institute and State University, Blacksburg, VA 24061}
\date{\today}

\begin{abstract}
The Mori-Zwanzig projection formalism is widely used in studying  systems with many degrees of freedom. We used a system-bath Hamiltonian system to show that the Mori's and Zwanzig's projection procedures are mutual limiting cases of each other depending on the size of the projected Hilbert space.  We also derived the dynamic equations of collective coordinates of a Hamiltonian system.
\end{abstract}

\pacs{Valid PACS appear here}
\maketitle

\section{\label{sec:Introduction}Introduction}
It is common to study dynamics of a system with  a large number of degrees of freedom in almost every scientific field. In general it is impractical, and often unnecessary,  to track all the dynamic information of the whole system. A common practice is projecting the dynamics of the whole system into that of a smaller subsystem through information contraction. The procedure leads to the celebrated Langevin and generalized Langevin dynamics. The Mori-Zwanzig formalism is a formal procedure of projection, especially for Hamiltonian systems. The original Mori procedure projects dynamics of the whole system into a sub-Hilbert space.  In the literature application of the Mori procedure generally results in a generalized Langevin equation(GLE)  that is "inherently linear in the system variables" \cite{Zwanzig2001}. The projection procedure developed by Zwanzig works on an enlarged Hilbert space, on the other hand, can lead to nonlinear GLEs \cite{Zwanzig1960,Zwanzig1961,Mori1965}. Recently Lange and Grubmuller tried to derive the dynamic equations of some collective coordinates with the Zwanzig projection procedure \cite{Lange2006}. 
Chorin and coworkers discussed generalizing the projection technique to non-Hamiltonian systems \cite{Chorin2000}.  In this communication we will show that the Mori's and Zwanzig's procedures are mutually limiting case to each other. We will also rederive the dynamic equations of collective coordinates, which differ slightly from that obtained by Lange and Grubmuller. 

\section{\label{sec:Theory}Theory}
First we will summarize the Mori procedure. We will follow the discussions given in \cite{Zwanzig2001} with some modifications, and focus on the Hamiltonian systems.

Consider a system described by the Hamiltonian,
\begin{eqnarray}
H(\mathbf{x,p}) = \sum_{i=1}^n\frac{p_i^2}{2}+V(\mathbf{x})
\end{eqnarray}
where $\mathbf{x}$ and $\mathbf{p}$ are position and conjugate momentum vectors. We will use mass-weighted coordinates throughout this paper.

The Liouville operator $L$ is defined as,
\begin{eqnarray}
LA = \sum_i \left( \frac{\partial H}{\partial p_i}\frac{\partial A}{\partial x_i} - \frac{\partial H}{\partial x_i}\frac{\partial A}{\partial p_i}  \right)
\end{eqnarray} 
For an arbitrary dynamic variable $A$, in this case the projection operator  is defined as,
\begin{eqnarray}
PA  &=&  \sum_{ij} (A, \phi_i)(\phi, \phi)^{-1}_{ij}\phi_j
\end{eqnarray}
$\{\phi(\mathbf{x},\mathbf{p} )\}$ composes the basis set for the projected subspace. The inner product  for two arbitrary variables $A$ and $B$ is defined as,
\begin{eqnarray}
(A,B) &=& <A^\dagger B>\nonumber\\
&=&\frac{\int   A^\dagger B \exp\left( -\beta H \right) d \mathbf{x} d \mathbf{p} }{\int  \exp\left( -\beta H \right) d \mathbf{x} d \mathbf{p} }
\end{eqnarray}
where $\dagger$ means taking transpose and complex conjugate. Any dynamic variable within the subspace can be expressed as a linear combination of the basis functions. The projected equations of an arbitrary dynamic variable $A$, which is defined within the projected subspace, are given by,
\begin{eqnarray}
\frac{\partial}{\partial t}A(t) = PLA(t) -\int_0^t ds\mathbf{K}(s)\cdot \phi(\mathbf{x}(t-s),\mathbf{p}(t-s) + F(t) \label{eqn:GLE_formal}
\end{eqnarray}
where 
\begin{eqnarray}
F(t)  &=&  \exp(t(\mathbf{1-P})L)(\mathbf{1-P})LA\\
\mathbf{K}(t) &=& -(LF(t),\phi) \cdot (\phi,\phi)^{-1} \nonumber\\
&=& (F(t),L\phi)\cdot (\phi,\phi)^{-1} \label{eqn:memory_general}
\end{eqnarray}
The last equation leads to the generalized fluctuation-dissipation relation, and we have used the anti-Hermitian property of the Liouville operator.

In practice the basis sets are usually chosen as portion of the coordinate vector $\mathbf{x}$   and the corresponding conjugate momentum vector $\mathbf{p}$.Then the Mori projection procedure results in a GLE that is linear to the coordinates and momentum. However, in principle this restriction is unnecessay. One can expand the Hilbert space
to include high order combinations of the coordinates and momentum, as shown by the example discussed below. Here we examine the extreme limit of including all the possible Hilbert functions composed by the coordinate and velocity (or momentum) in reduced dimension.  The following procedure is analogous to what adopted by Zwanzig \cite{Zwanzig1961}. 
For simplicity let's focus on a one-dimensional coordinate and its conjugate momentum, while generalization to higher dimensions is straightforward,
 \begin{eqnarray}
 c &=& f(\mathbf{x})= f(0) + \nabla_{\mathbf{x}}f(0)\cdot \mathbf{x} + \frac{1}{2}\nabla_{\mathbf{xx}}f(0)\colon \mathbf{xx} +\dots\\
 \dot{c} &=&  \nabla_{\mathbf{x}} f \cdot \mathbf{\dot{x}}= \nabla_{\mathbf{x}} f(0) \cdot \mathbf{\dot{x}}
    + \nabla_{\mathbf{xx}} f(0) \colon \mathbf{x\dot{x}}+\dots
 \end{eqnarray}
Therefore $c$ and $\dot{c}$ are vectors in the full Hilbert space. Let's consider the sub-Hilbert space, which may still have infinite dimension, supported by  all the possible multiplicative combinations of $c$ and $\dot{c}$ , such as $c^2, \dot{c} c^3\dots$. 
 A key observation is that these basis functions, denoted $\{\phi(c,\dot{c})\}$, compose a complete basis set for the subspace so any arbitrary function of $(c,\dot{c})$,  can be fully expressed by the basis set. That is, for an arbitrary function $g(\mathbf{x,p})$,
 \begin{eqnarray}
 \sum_i\int g \phi_i \exp(-\beta H) d\mathbf{x}d\mathbf{p} \nonumber\\
 =\frac{1}{ \bar{\rho}(c,\dot{c})}
  \int g  \exp(-\beta H) \delta(f-c)\delta(\nabla_\mathbf{x} f \cdot \mathbf{p} - \dot{c}) d\mathbf{x}d\mathbf{p} 
 \end{eqnarray} 
 where
 \begin{eqnarray}
 \bar{\rho}(c,\dot{c}) = {\int   \exp(-\beta H) \delta(f-c)\delta(\nabla_\mathbf{x}f \cdot \mathbf{p} - \dot{c}) d\mathbf{x}d\mathbf{p} }
 \end{eqnarray}
The above expression may be more familiar if the Dirac bra and ket notations are used. Then for the projected equations, 
\begin{eqnarray}
PL \cdot c &=& \dot{c} \label{eqn:collective_c}\\
PL \cdot \dot{c} &=&  -k_BT\frac{1}{\bar{\rho}(c,\dot{c})}\frac{\partial}{\partial_c}  \int   \exp(-\beta H) \nonumber\\
      && ||\nabla_{\mathbf{x}} f ||^2 \delta(f-c)\delta(\nabla_\mathbf{x}f \cdot \mathbf{p} - \dot{c}) d\mathbf{x}d\mathbf{p}\label{eqn:collective_cdot} 
\end{eqnarray} 
To derive the above expression, we performed integration by parts, and used the relations,
\begin{eqnarray}
\nabla_{\mathbf{x}} \delta(c-f) &=&  \nabla_{\mathbf{x}}f \partial_f \delta(x-f) =   \nabla_{\mathbf{x}}f \partial_c\delta(x-f) \nonumber\\
\nabla_{\mathbf{x}} \delta(\dot{c}-\nabla_{\mathbf{x}}f \cdot \mathbf{p}) &=& \nabla_{\mathbf{x}} (\nabla_{\mathbf{x}}f \cdot \mathbf{p}) \partial_{\dot{c}} \delta(\dot{c}-\nabla_{\mathbf{x}}f \cdot \mathbf{p})  \nonumber\\
\nabla_{\mathbf{p}} \delta(\dot{c}-\nabla_{\mathbf{x}}f \cdot \mathbf{p}) &=& \nabla_{\mathbf{x}}f  \partial_{\dot{c}} \delta(\dot{c}-\nabla_{\mathbf{x}}f \cdot \mathbf{p})
\end{eqnarray}
Compared to Eq. \ref{eqn:collective_cdot}, the result derived by Lange and Grubmuller has an extra term $||\nabla_{\mathbf{x}}f||^2$ in the expression of $\bar{\rho}(c, \dot{c} )$ \cite{Lange2006}. The discrepancy may come from the fact that the projection operator
defined in \cite{Lange2006} does not rigorously satisfy $P^2=P$. It remains to be examined on how this extra term may affect the dynamics. In the case $f$ is a linear combination of $\mathbf{x}$, and is chosen to satisfy $||\nabla_{\mathbf{x}}f||^2=1$ , Eq. \ref{eqn:collective_cdot}
gives the familiar relation to the gradient of potential of mean force.  

\section{\label{sec:Example}Example}
Here we consider a system-bath Hamiltonian,
\begin{eqnarray}
H=\frac{1}{2}p^2+\frac{1}{2}x^2+\frac{b}{4}x^4 + \sum_j\left\{\frac{1}{2}p_j^2+\frac{1}{2}\omega_j^2\left(q_j-\frac{\gamma_j}{\omega_j^2 }x \right)^2\right\}
\end{eqnarray}
Zwanzig discussed a nonlinear GLE for the system coordinates $\{x, p\}$  obtained by directly solving the equations of motion \cite{Zwanzig1973, Zwanzig2001}, 
\begin{eqnarray}
\frac{dx(t)}{dt} &=& p(t)\nonumber\\
\frac{dp(t)}{dt} &=& -x(t)-bx(t)^3 \nonumber\\
&& -\int_0^t dsK_N(s)p(t-s)+F_N(t)\label{eqn:GLE_exact}
\end{eqnarray}
The memory kernel and the random force terms are given by,
\begin{eqnarray}
K_N(t) &=& \sum_j \frac{\gamma_j^2}{\omega_j^2}\cos(\omega_j t) \label{eqn:memory_exact}\\
F_N(t)  &=& \sum_j \gamma_j p_j(0) \frac{\sin \omega_j t} {\omega_j} \nonumber\\
&& + \sum_j \gamma_j \left(q_j(0)-\frac{\gamma_j}{\omega_j^2}x(0)\right)\cos \omega_j t\label{eqn:randforce_exact}
\end{eqnarray}
with the fluctuation-dissipation relation,
\begin{eqnarray}
<F_N(t) F_N(t')>_0 &=& k_BTK_N(t-t').
\end{eqnarray}
The average is over an equilibrium heat bath with the system constrained at $\{x(0),p(0)\}$.
By projecting to the Hilbert space $(x,v)$ with Mori's procedure, one can also obtain a linearized GLE \cite{Zwanzig2001},
 \begin{eqnarray}
\frac{dx(t)}{dt} &=& v(t)\nonumber\\
\frac{dp(t)}{dt} &=& -\omega_0^2 x(t)-\int_0^t dsK_L(s)p(t-s)+F_L(t)
\end{eqnarray}
Where $\omega_0^2=k_BT/<x^2>$, and the the random force and memory kernel terms are also related by the fluctuation-dissipation relation
\begin{eqnarray}
<F_L(t) F_L(t')> &=& k_BTK_L(t-t'). 
\end{eqnarray}
However in this case, the average is over the unconstrained thermal equilibrium distribution. Effects of the nonlinear term $-bx(t)^3$ are contained in the renormalized
coefficent $\omega_0^2$, the memory kernel, and the random force terms.

In the following discussions, we will generalize the projection procedure of Mori by choosing a basis set $\{x,x^3,v\}$. Functions 
with even powers of $x$ makes no contribution to the projection ($(Lp,x^{2n})=0, n=1,2,\dots$). Therefore the lowest nolinear basis function is $x^3$.

First,
\begin{eqnarray}
Lx = p, Lx^3 = 3px^2\nonumber\\
Lp = -x-bx^3-\sum_i \gamma_i\left(\frac{\gamma_i}{\omega_i^2}x-q_i \right)
\end{eqnarray}
 let's calculate the normalization matrix,
\begin{eqnarray}
A^{-1}&=& \left(\begin{array} {ccc}
          <x^2> & <x^4>   & <xp>\\
          <x^4> & <x^6>   & <x^3p>\\
          <xp>   & <x^3p> & <p^2>
          \end{array}    
\right)^{-1}\\
&=& \left(\begin{array} {ccc}
          <x^6>/h & -<x^4>/h   & 0\\
          -<x^4>/h & <x^2>/h   & 0\\
          0  & 0 & <p^2>^{-1}
          \end{array}\right)    
\end{eqnarray}
Where $h=<x^2><x^6>-<x^4>^2$.
The memory function and the random force in the equation of motion of $x$ vanish, which can be seen from,
\begin{eqnarray}
Lx &=&  \left(\begin{array} {ccc}
          (Lx,x) & (Lx,x^3)   & (Lx,p)
          \end{array}\right)\cdot A^{-1} \nonumber\\
          &=& p   
\end{eqnarray}
One has,
\begin{eqnarray}
(Lp,x^n) &=& -\frac{1}{\int \exp(-\beta H)dx}\int x^n \frac{\partial H }{\partial x} \exp(-\beta H)dx\nonumber\\
&=& \frac{k_BT}{\int \exp(-\beta H)dx}\int x^n \frac{\partial}{\partial x} \exp(-\beta H)dx\nonumber\\
&=& -\frac{k_BT}{\int \exp(-\beta H)dx}\int nx^{n-1} \exp(-\beta H)dx\nonumber\\
&=&   -n k_BT<x^{n-1}>\\
(Lp,x) &=& -k_BT= -\omega_0^2<x^2>
\end{eqnarray}
However, one also has,
\begin{eqnarray}
<Lp,x^n> &=& -<x^{n+1}> -b<x^{n+3}>  \nonumber\\
&&- \sum_i \gamma_i <x^n (\frac{\gamma_i}{\omega_i^2}x - q_i)>\nonumber\\
&=& -<x^{n+1}> -b<x^{n+3}> 
\end{eqnarray}
Therefore,
\begin{eqnarray}
<x^4> &=& \frac{1}{b}(\omega_0^2-1)<x^2>\\
<x^6>  &=&  \frac{3}{b} \omega_0^2<x^2><x^2> \nonumber\\
&& -\frac{1}{b^2}(\omega_0^2-1)<x^2>
\end{eqnarray}
  Then,
\begin{eqnarray}
 PL p(t) = -x-bx^3
\end{eqnarray}
One can easily show that the random force (through $dF/dt = (1-P)LF$) and memory kernel (through Eq. \ref{eqn:memory_general}) terms are the same as those given in Eqs. \ref{eqn:memory_exact} and \ref{eqn:randforce_exact} , although in general here the average perform in Eq. \ref{eqn:memory_general} is over the unconstrained thermal equilibrium distribution. Therefore with the Mori projection procedure we recover Eqs. \ref{eqn:GLE_exact}, \ref{eqn:memory_exact}, \ref{eqn:randforce_exact}  obtained by exact integration.  Following similar procedure, one can show that further expanding the basis functions to include higher orders of $x^n$ does not change the projected equation form.
The above results can also be obtained by applying Eqs. \ref{eqn:GLE_formal}, \ref{eqn:collective_c},\ref{eqn:collective_cdot} directly.


\begin{references}
\bibitem{Zwanzig2001} R. Zwanzig, Nonequilibrium Statistical Mechanics (Oxford University Press, Oxford, 2001).  
\bibitem{Zwanzig1960} R. Zwanzig, J. Chem. Phys. 33, 1338 (1960).
\bibitem{Zwanzig1961}R. Zwanzig, Phys. Rev. 124, 983 (1961).
\bibitem{Mori1965} H. Mori, Prog. Theor. Phys. 33, 423 (1965).
\bibitem{Lange2006} O. F. Lange and H. Grubmuller, J. Chem. Phys. 124, 214903 (2006).
\bibitem{Chorin2000} A. J. Chorin, O. H. Hald, and R. Kupferman, Proc. Natl. Acad. Sci. USA 97, 2968 (2000).
\bibitem{Zwanzig1973} R. Zwanzig, J. Stat. Phys. 9, 215 (1973).
\end{references}

\end{document}